# Gate-Tuned Spontaneous Exciton Insulator in Double-Quantum Wells

Lingjie Du[1], Wenkai Lou[2], Kai Chang[2], Gerard Sullivan[3], and Rui-Rui Du[1,4]*

[1]Department of Physics and Astronomy, Rice University, Houston, Texas 77251-1892, USA.
[2]SKLSM, Institute of Semiconductors, Chinese Academy of Sciences, Beijing 100083, China
[3]Teledyne Scientific and Imaging, Thousand Oaks, CA 91630, USA.
[4]International Center for Quantum Materials, Peking University, Beijing 100871, China
*Correspondence to:rrd@rice.edu(R. R. Du)

(Dated: July 27, 2015)

## Abstract

It was proposed that a dilute semimetal is unstable against the formation of an exciton insulator, however experimental confirmations have remained elusive. We investigate the origin of bulk energy gap in inverted InAs/GaSb quantum wells (QWs) which naturally host spatially-separated electrons and holes, using charge-neutral point density ($n_o \sim p_o$) in gated-device as a tuning parameter. We find two distinct regimes of gap formation, that for I), $n_o \gg 5\times10^{10}/cm^2$, a soft gap opens predominately by electron-hole hybridization; and for II), approaching the dilute limit $n_o \sim 5\times10^{10}/cm^2$, a hard gap opens leading to a true bulk insulator with quantized edge states. Moreover, the gap is dramatically reduced as the QWs are tuned to less dilute. We further examine the response of gaps to in-plane magnetic fields, and find that for I) the gap closes at $B_{//} > \sim 10T$, consistent with hybridization while for II) the gap opens continuously for $B_{//}$ as high as 35T. Our analyses show that the hard gap in II) cannot be explained by single-particle hybridization. The data are remarkably consistent with the formation of a nontrivial exciton insulator in very dilute InAs/GaSb QWs.

**Introduction**

Exciton insulator (EI) was proposed several decades earlier (1-4) to be an equilibrium ground state in certain dilute semimetals. In semimetals, with negative binding energy exciton state forms spontaneously, giving a gapped spectrum without optical pumping. Later it was proposed that exciton condensation and superfluidity could be achieved in double-layer systems with spatially-separated electrons and holes, see *e. g.*, Ref. *5*. While such macroscopically coherent quantum states have been observed in quantum Hall double-layers (*6*), the evidence for their existence in a two-band electron-hole double layer and in the absence of a quantizing magnetic field is still lacking. Recently the topological classification(7,8) of EIs(9-12) is considered with proposed topological exciton insulator (TEI)(13-15). Realization of nontrivial exciton bulk state is a crucial step in exploring exotic phenomena of interacting topological insulators.

It was recently observed that Quantum spin Hall (QSH)(*16*) effect in inverted InAs/GaSb quantum wells(QWs) is extremely robust against the temperature and magnetic field(*17*), with theory proposing(*14*) that TEI emerges in the bulk and accounts for such unexpected properties. Inverted InAs/GaSb QWs(Fig. 1A) has negative band gap allowing coexistence of spatially-separated electrons and holes, offering a natural setting for EI(*10*). However, to date there has been no direct experimental evidences for an exciton ground state in the QWs. Major challenge is to achieve the dilute limit, where electron and hole will bind to form exciton instead of collective e-h plasma(Fig. 1B). Moreover, the interlayer tunneling which is the mechanism for forming hybridization gap, should be properly controlled because it mixes bands thereby reducing binding energy as well as exciton lifetime (*9*).

We report experimental evidences for EI gap in double-layer InAs/GaSb QWs. Using a pair of gates, we are able to tune the electron and hole densities to the dilute limit $4n_o a_o^2 \sim 1$ where a truly insulating bulk and quantized edge states are observed ($n_o$ is equilibrium density and $2a_o$ is Bohr diameter, see Fig. 1A). Characteristically, the gap remains open in an in-plane magnetic field up to 35T, where interlayer tunneling is eliminated. This result confirms that the dilute-limit gap is originated from excitons, rather than hybridization. In sharp contrast, in the opposite limit, $4n_o a_o^2 \gg 1$, we observe gaps that are vanishing above 10T, consistent with a hybridization origin.

**Tuning into dilute limit of carrier densities**

Our devices are made from inverted InAs/GaSb QWs (Wafer A and B) grown by MBE(*17*); both wafers yield very similar results so most data shown here are from A, except for those otherwise indicated. The electron density in InAs and the hole density in GaSb can be tuned by front and back-gates. As shown in Fig. 1C, with both gates the charge-neutral point (CNP) density $n_0 = p_0$ can be gated continuously in the range 5-9 $\times 10^{10}$cm$^{-2}$. Fig. 1D and 1E display $B/eR_{xy}$ vs $\Delta V_f$ for a Hall bar device at $V_b$=-6V (case I, less dilute) and 0V (case II, dilute limit), respectively; $B$ is the perpendicular magnetic field, $\Delta V_f$ is the bias increment from the CNP. At high electron density, $B/eR_{xy}$ is consistent with the density obtained from the period of Subinikov de Haas oscillations. For $V_b$=-6V, as more holes are introduced by backgate, two-carriers transport dominates and e-h hybridized regime is reached, as indicated by divergent $R_{xy}$ traces around CNP(16, 18). For $V_b$=0V, as CNP is approached, a plateau-like feature in $B/eR_{xy}$ is observed. This is due to non-local transport in the asymmetrical Hall bar with insulating bulk and helical edge. Note that here the dilute limit corresponds to a very shallow inversion, just above the TI - normal insulator transition.

In the dilute limit, to further explore the insulating bulk at CNP, we measure the conductance $\sigma_{xx}$ in a double-gate Corbino device C1(Fig. 2A). Since the edge states are shunted by concentric metallic contacts, $\sigma_{xx}$ exclusively measures the bulk properties. As shown in Fig. 2B, for $n_o = 9 \times 10^{10}$cm$^{-2}$ and under 0T there is a $\sigma_{xx}$ dip (red trace) around CNP which corresponds to hybridization gap. In subsequent panels from left to right corresponding to decreasing $n_o$, the $\sigma_{xx}$ dip goes down to zero; the width of zero conductance around CNP increases. These data obtained at zero B have already shown a "soft gap" (*i. e.*, finite density of states in the gap) to a "hard gap" transition tuned by gates.

**Suppressing interlayer tunneling by magnetic barrier**

To delineate the origin of the soft gap and that of the hard gap, we apply an in-plane magnetic field $B_{//}$ to the same Corbino device and measure the $\sigma_{xx}$ as a function of $B_{//}$. Within the single-particle hybridization picture, electrons and holes with the same Fermi momentum $k^e_F$ and $k^h_F$ tunnel between two QWs, forming mini-gaps(*19, 20*). Assuming that $B_{//}$ applies along *y* axis in the plane, Lorenz force gives tunneling carriers additional momentum along *x* axis, resulting in a relative shift of band dispersions $\Delta k_x=-eB\Delta<z>/h$, (tunneling distance $\Delta<z>$ is limited by one-half thickness of QWs). Consequently, the inter-well tunneling is suppressed due to momentum-mismatch. This effect from $B_{//}$ is also referred to as "magnetic barrier" (*21*), indicating that $B_{//}$ effectively creates additional potential barrier for tunneling. Therefore elastic tunneling is forbidden in very high fields. As shown by 8-band self-consistent calculation using our device parameters(Fig. 2C), as soon as $B_{//}$ increases beyond 18T two bands are separated in momentum space and the system becomes an indirect double-layer semimetal. In this case there exist no tunneling channels at low temperatures.

With this picture in mind we inspect the panels in Fig. 2B, but now paying attention to the $\sigma_{xx}$ traces under 35T (blue). We immediately notice that the response of $\sigma_{xx}$ to $B_{//}$ dramatically depends on $n_o$. For $n_o = 9 \times 10^{10} cm^{-2}$, $\sigma_{xx}$ increases from the B = 0 value by 4 fold. For $n_o = 5.6 \times 10^{10} cm^{-2}$, $\sigma_{xx}$ remains zero but its width decreases. Finally, for the lowest density $n_o = 5 \times 10^{10} cm^{-2}$, $\sigma_{xx}$ is characteristically the same as that of B = 0. Together with the $\sigma_{xx}$ results at zero field (red), these observations can be interpreted as what follows: the departure of $\sigma_{xx}$(35T) from $\sigma_{xx}$(0T) is a qualitative measure of how much contributions from single-particle hybridization to the gap formation, using $n_o$ as a tuning parameter. We can naturally infer that at the *lowest density* hybridization does not play much role on the gap formation. On the other hand, as more evidences will be shown later, spontaneous exciton binding emerges as a leading mechanism for forming the gap, and hence the exciton insulator becomes the ground state of dilute double QWs.

**Gap energy and edge states in dilute limit**

From now on we concentrate on the dilute limit, *i.e.*, $n_o = 5 \times 10^{10} cm^{-2}$. Fig. 3A shows a plot of $\sigma_{xx}$ in a series of $V_f$-sweeps under consecutive and fixed $B_{//}$, where a broad zero conductance (between ±0.5V) can be seen from $B_{//}$= 0 to 35 T. In other words, the gap remains open continuously in spite of strong external magnetic fields. The same result is obtained for Corbino device C2 made from Wafer B. That the gap remains open is further confirmed *quantitatively* by thermal activation energy measurements. Shown in Fig. 3C as an example (for device C1 at $B_{//}$ = 35T), the $\sigma_{xx}$ vs. *1/T* can be fitted over two order of magnitude into an Arrhenius plot $\sigma_{xx} \propto exp(-\Delta/2kT)$, with gap energy $\Delta$ = 25K. Following the same procedures we obtain $\Delta = 27 \pm 1$K for $B_{//} = 0$, 9, 18, 27, and 35T, with a peak value 28K near 9T.

Remarkably, our experiments have also confirmed the helical edge transport throughout the entire field range. To this end, nonlocal measurement(*22*) in the dilute limit is performed in mesoscopic H bar device H1. The electrical current is passed through contacts 3 and 4 and the voltage is measured between contacts 1 and 2. In an ideally bulk-insulating QSH insulator, the currents go through all contacts with the path surrounding the bulk. According to Landauer-Büttiker formula(*23*), $R_{12,34}$ should measure quantized resistance of $h/4e^2$ ~ 6.45kΩ. Under both $B_{//}$= 0T and 35T, as shown in Fig. 3E and 3F, we indeed observe quantized plateau close to this value. The same observations are true for device H2 from wafer B, as shown in Fig. 3G and 3H. Nonlocal results not only confirm the bulk of the dilute-limit QWs is truly insulating, but should also be taken as direct evidence for a topological insulator originated from excitonic ground state.

**Formation of an excitonic ground state in the dilute limit**

As proposed in Ref. (1-5, 9, 10), a dilute semimetal system in spatially-separated QWs is unstable against the formation of a spontaneous exciton ground state driven by Coulomb and exchange interactions. Here the diluteness of the CNP carriers is a crucial parameter for semimetal to exciton transition, because in the opposite limit more free carriers will participate in screening process thereby reducing exciton binding energy. Moreover, in our QWs without a potential barrier, certain interlayer tunneling exists which would mix the bands, hence also reduce the binding energy(*9*). However, with $B_{//}$ = 35T the tunneling phase space vanishes, together with a low density $n_o = 5 \times 10^{10}$cm$^{-2}$, creating a set of more favorable conditions for forming an exciton gap. This point is reinforced by the fact that gap $\Delta = 25K$ agrees with the theoretical calculation of EI binding energy(10-12), but far away from hybridization gap energy ~ 50K. Moreover, according to EI model(*10*) as well as measured $\Delta$ and $n_o$, we obtain Bohr diameter $2a_o=2\hbar/(2m\Delta)^{0.5}=49$nm, where m is the exciton reduced mass. With intra-layer carrier distance 45nm in dilute limit which is comparable with Bohr diameter, the bulk satisfies critical condition $4n_o a_o^2 \sim 1$ for spontaneous excitons. This crude estimate suggests that the system is entering the dilute limit for EI formation.

Returning to zero magnetic field (Fig. 4), we measure the gap as a function of $n_o$ for a quantitative confirmation of "soft gap" to "hard gap" transition when approaching the lowest density. The data at CNP follows well the relation $\sigma_{xx} \propto exp(-\Delta/2kT)$ yielding a set of $\Delta$ as plotted in Fig. 4B. Most interestingly the gap energy diminishes steeply to nearly zero as $n_o$ increases by just a factor of 2(Fig. 4C). Hence, the gap energy is strongly correlated with $1/n_o$, agreeing with the properties of EI. Note that with increasing $n_o$, the Bohr radius increases reflecting the free carrier screening effect. As $n_o$ increases, $n_o a_o^2$ dramatically increases from ~ 0.25 to 10, suggesting de-binding of exciton states (Fig. 4D).

**Discussion and conclusion**

While most proposals for bilayer exciton insulator uses a thin layer between two QWs as a barrier, an exciton ground states in the present structure (*i.e.,* without an explicit barrier) is still possible, as has been theoretically addressed in Ref. (10, 14). In particular, Ref. *14* proposes that in InAs/GaSb QSH effect the interplay of spin and exciton condensation would lead to a rich phase diagram for exciton order parameters, including the *s-wave* (trivial) to *p-wave* (non-trivial) phase transition driven by interlayer tunneling. Such transition could be accessed by future experiments on QWs with various barrier layers. Overall, the data here in the dilute-limit suggest that the zero-field state and 35T state are adiabatically connected, showing no experimental sign of topological transition.

We comment on the relation of external magnetic field and time reversal symmetry (TRS) protection in this system, under different carrier diluteness. In the regime I (hybridization) an in-plane field shifts momentum, and $k_x$ and $-k_x$ are not equivalent hence

TRS is explicitly broken (referring to Fig.2C). On the contrary, in the regime II (exciton) the $k_x$ and $-k_x$ remain equivalent regardless of an external magnetic field. In other words here the robustness of the QSH insulator may be understood from the view point of TRS protected topological phase.

In conclusion, we report on a systematic study of inverted InAs/GaSb double-QWs, and the data are remarkably consistent with the formation of a topological exciton insulator as the CNP density approaches dilute limit. Our findings provide compelling experimental observation of spontaneous exciton ground state. Future work can be expected for bilayer superfluidity and Bose-Einstein condensation, and others, in this highly tunable e-h system.

**Acknowledgments**

We acknowledge conversations with A. H. MacDonald, B. I. Halperin, and D. I. Pikulin. L.J.D. was supported by DOE Grant (No.DE-FG02-06ER46274). R.R.D. was supported by NSF Grants (No. DMR-1207562 and DMR-1508644). W.K.L. and K.C. were supported by NSFC (No. 11434010). The work in PKU was supported by NBRPC Grant (No. 2014CB920901). A portion of this work was performed at the National High Magnetic Field Laboratory, which is supported by NSF Cooperative Agreement under No. DMR-1157490, and by the State of Florida.

**Fig. 1.** Device layout and density measurement under double-gate control. (**A**)Sketch of device layout. Holes are in GaSb layer (green) while electrons are in InAs layer (red). Front and back gate(blue) are fabricated to tune carrier densities. (**B**)Phase diagram of exciton with temperature and $n_o a_o^2$. (**C**)shows CNP density $n_o$ as a function of $V_f$ and $V_b$ in units of $10^{10}$/cm$^2$ . (**D**) and (**E**) are $B/eR_{xy}$ vs $V_f$ traces of the asymmetric 50μm x 50μm Hall bar for $V_b$= -6V and 0V, respectively. Data are taken at 300mk with 1T perpendicular magnetic field. The inset in (**D**) is a schematic of the asymmetric Hall bar. The region in the dashed box is covered by front gate. Insets in (**D**) and (**E**) are showing band alignments corresponding to the deeply- and shallowly-inverted regime, or dense and dilute $n_o$, respectively.

**Fig. 2.** Corbino measurement under an in-plane magnetic field. (**A**). Sketch of measurement setup. With outer conductor grounded, a voltage is applied in the inner conductor and the current flowing across the ring is measured. $V_f$ and $V_b$ are applied to modulate the carriers in the ring. (**B**) Conductance measured at T = 30 mK. The plots show $V_f$ dependence of the conductance $\sigma_{xx}$ for device C1 from $V_b$ = -6V to 0V, with the decrement of 1.5V. $n_o$ is noted in units of $10^{10}$/cm$^2$. The blue lines are for $B_{//}$=35T while the red lines are for 0T. As $|V_b|$ decreases, $n_o$ decreases and an insulator emerges. (**C**). The energy dispersion calculated from the 8-band self-consistant model for tunneling electrons and holes under $B_{//}$=0T, 9T, 18T and 35T, respectively.

**Fig. 3.** Gap measurement and nonlocal measurement under $B_{//}$=35T. (**A**) shows $V_f$ dependence of the conductance $\sigma_{xx}$ for device C1 in dilute limit under $B_{//}$ from 0T up to 35T. Traces are taken at 30mK. Zero-conductance dip appears and persists up to 35T. The color corresponds to $\sigma_{xx}$ in units of Siemens. (**B**) shows the conductance traces under temperature (5.8K, 5K, 4.5K, 3.95K, 3.45K, 2.95K, 2.6K and 2.29K) in dilute limit and under 35T. (**C**). Dependence of dip conductance in Corbino measurement on 1/T under 35T. Solid line is guide to the eye. The data can be fitted by $\sigma_{xx} \propto exp(-\Delta/2kT)$ to obtain $\Delta$. (**D**). Gap energy $\Delta$ obtained for a series of $B_{//}$ in dilute limit. (**E**) and (**F**) show nonlocal measurement performed in meso-H bar from wafer A under 0T and 35T, respectively. The dotted lines indicate the expected resistance value from Landauer-Büttiker formula. (**G**) and (**H**) show nonlocal measurement performed in meso-H bar from wafer B under 0T and 35T, respectively. In the gap, the current path is shown in the inset of (**E**) as red and green arrows.

**Fig. 4.** Measurements of the $n_o$-dependent gap energy $\Delta$ under B = 0T. (**A**) shows the conductance $\sigma_{xx}$ traces measured at various fixed temperatures (5K, 4.6K, 4K, 3.5K, 3K, and 2.6K) (**B**). Dependence of the conductance minimum on 1/T for different $V_b$. Here the $\sigma_{xx}$ is normalized by $\sigma_{xx}= \sigma_{xx\ min}/ (\sigma_{xx\ min}$ at ~ 2.5K). Solid lines are guides to the eye. Black circles are for $V_b$=0V, red squares are for $V_b$=-1.5V, green down-triangles are for $V_b$=-3V, blue up-triangles are for $V_b$=-4.5V, and yellow stars are for $V_b$=-6V. (**C**) shows the measured gap energy $\Delta$ as a function of $n_o$. (**D**) shows $n_o$ dependence of $n_o a_o^2$. At the lowest density, $n_o a_o^2$ is approximately 0.25.

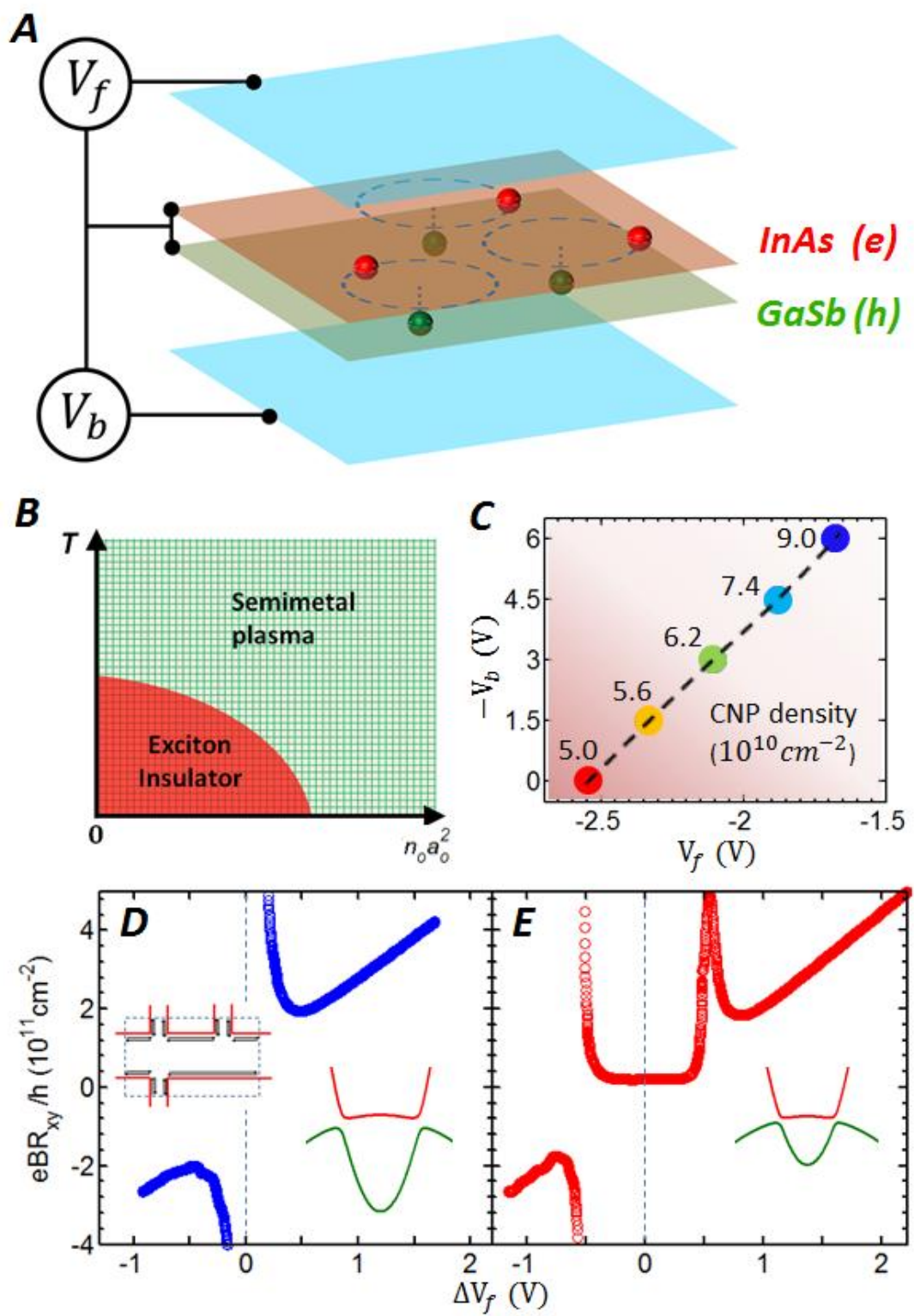

**Fig. 1**

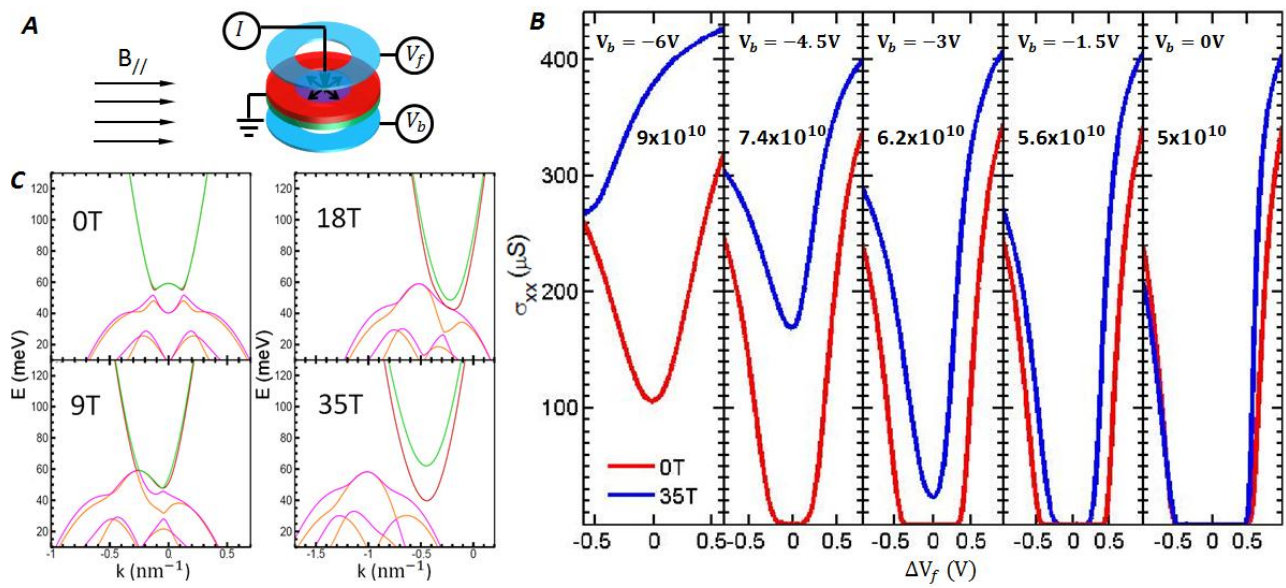

**Fig. 2**

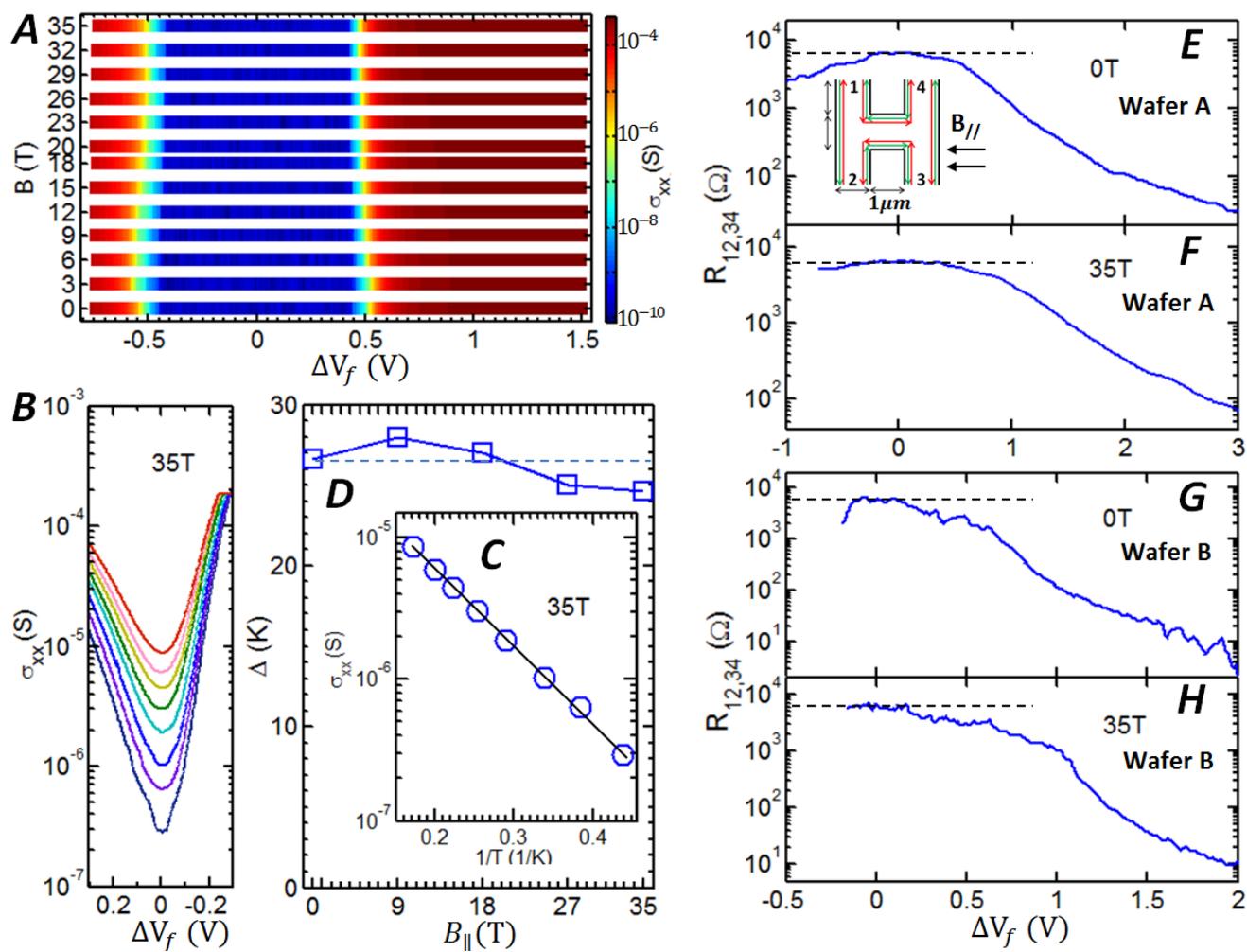

**Fig. 3**

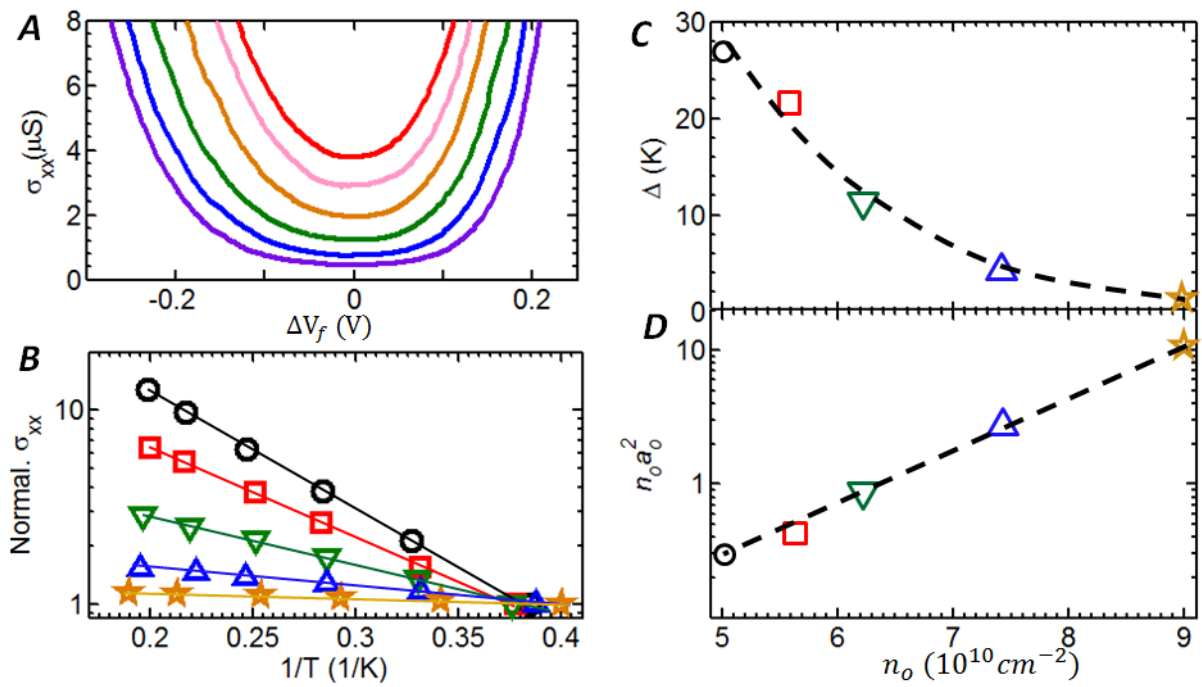

**Fig. 4**